\documentclass[twoside,11pt]{article}

\usepackage{multirow}
\usepackage{xcolor}
\usepackage{amsmath}
\usepackage{amssymb}
\usepackage{amsbsy}
\usepackage{epsfig}
\usepackage{color}
\usepackage{url}
\usepackage{setspace}
\usepackage{color}
\usepackage{bm}
\usepackage{graphicx}
\usepackage{rotating}
\usepackage{dsfont}
\usepackage{float}
\usepackage{algorithm}
\usepackage{algpseudocode}
\usepackage{xcolor}
\usepackage{mathrsfs}
\usepackage{subcaption}
\usepackage{natbib}
\usepackage{dsfont}
\usepackage[margin=1in]{geometry}
\newcommand{\bell}{\bm{\ell}}
\newcommand{\bu}{\bm{u}}
\renewcommand{\tilde}[1]{\widetilde{#1}}
\algnewcommand\algorithmicforeach{\textbf{for each}}
\algdef{S}[FOR]{ForEach}[1]{\algorithmicforeach\ #1\ \algorithmicdo}
\renewcommand{\hat}[1]{\widehat{#1}}

\begin{document}

\title{Adjusting for Spatial Correlation in Machine and Deep Learning}

\author{Matthew~J.~Heaton \\ mheaton@stat.byu.edu \\
       Department of Statistics\\
       Brigham Young University\\
       Provo, UT 84602, USA \and
       Andrew~Millane \\ amillane@student.byu.edu \\
       Department of Statistics \\
       Brigham Young University \\
       Provo, UT 84602, USA \and
       Jake~S.~Rhodes \\ rhodes@stat.byu.edu \\
       Department of Statistics\\
       Brigham Young University\\
       Provo, UT 84602, USA
       }

\maketitle

\begin{abstract}%   <- trailing '%' for backward compatibility of .sty file
Spatial data display correlation between observations collected at neighboring locations.  Generally, machine and deep learning methods either do not account for this correlation or do so indirectly through correlated features and thereby forfeit predictive accuracy.  To remedy this shortcoming, we propose preprocessing the data using a spatial decorrelation transform derived from properties of a multivariate Gaussian distribution and Vecchia approximations. The transformed data can then be ported into a machine or deep learning tool. After model fitting on the transformed data, the output can be spatially re-correlated via the corresponding inverse transformation.  We show that including this spatial adjustment results in higher predictive accuracy on simulated and real spatial datasets.

\end{abstract}

\section{Research Synopsis and Goals}
``Spatial data,'' broadly speaking, is any data that contains geographic or location-based information about the observations. Spatial data are found in nearly every scientific field from ecology \citep{plant2018spatial} to environmental science, \citep{harris2014statistics,yuan2020deep}, public health \citep{waller2004applied,shaddick2015spatio}, real estate \citep{pace1998spatial, arbia2021spatial}, civil engineering \citep{ziakopoulos2020review} and beyond.  As a result, appropriate analysis of spatial data sets of increasing size and scope can transform a wide range of scientific disciplines.

The primary challenge of appropriately predicting spatial data lies in the inherent correlation arising between observations due to the geographic nature of the collected data. The field of spatial statistics is concerned with the development of such methods \citep{stein2022impact}.  The Gaussian process (GP) stands firmly at the forefront \citep{gelfand2016spatial} of spatial methodology because it directly accounts for the spatial correlation in the data.  While the flexibility of GPs allows them to be applied to many spatial problems, their lack of scalability to large datasets is an issue for modern spatial applications \citep{bradley2016comparison,heaton2019case,huang2021competition}.  Specifically, factorization of the covariance matrix is $\mathcal{O}(n^3)$ complexity.  Further, the GP typically relies on oversimplifying assumptions such as stationarity or linearity \citep{banerjee2014hierarchical} which may not hold in the data.

Because of their ability to capture non-linear relationships and computational scalability, recent attention has shifted to machine and deep learning (ML and DL) approaches for predicting with spatial data. 
 Recent spatial ML and DL approaches to spatial prediction and modeling include random forests \citep{nikparvar2021machine, patelli2023path}, fully connected neural networks \citep{wikle2023statistical}, compositional neural networks \citep{zammit2022deep}, graph neural networks \citep{sainsbury2023neural, cisneros2024deep, tonks2024forecasting} or even deep Gaussian processes \citep{vu2022modeling, sauer2023active}. Many of these implementation of ML or DL to spatial data involve engineering spatial features as inputs into the model architecture \citep{patelli2023path}.  For example, \citet{sekulic2020random} and \citet{georganos2021geographical} use spatial coordinates and spatial neighbors as inputs to ML algorithms.  \citet{gray2022use}, \citet{lin2023some} and \citet{chen2020deepkriging} use spatial basis functions \citep[see][]{bradley2011selection} as features in fully connected neural network models in a method they term ``deep kriging.''  \citet{zammit2024spatial} take this idea one step further allowing weights and biases of neural networks to vary over space.  \citet{} and \citet{} use graphical neural networks to perform kriging.

The issue with these many of these ML and DL approaches lies in their assumption that the inclusion of spatial features allows the model to adequately capture the spatial structure present in the data so that observations can be treated as independent in the loss function.  That is, each of these approaches typically relies on loss functions that are independent summations across the observations (i.e.\ $\sum_{i=1}^n(y_i-\hat{y}_i)^2$).  However, work by \citet{stein2014limitations} showed that spatial inputs likely oversmooth the data leading to residual spatial correlation.  To this end, \citet{saha2023random} and \citet{zhan2023neural} build ML and DL models which directly incorporate spatial correlation from the observations into the loss function by using a generalized squared error loss during the model training.  Using these generalized loss functions, however, can come at the cost of computational complexity unless special care is taken in their implementation via either computationally simplifying assumptions (e.g.\ Vecchia approximations, \citealt{katzfuss2021general}) or supercomputing \citep{abdulah2018exageostat}

The primary contribution of this research is to propose the use of a spatially decorrelating transformation as a preprocessing step when fitting ML and DL models on spatial data.  Using the form of the conditional distribution of a multivariate Gaussian distribution, our proposed transformation decorrelates the data by subtracting off the effect of spatial neighbors weighted by a parametric correlation function.  After applying this preprocessing step, the resulting decorrelated, transformed data can be input into the desired ML or DL model and trained with any standard loss functions rather than more computationally intensive generalized loss functions.  The output of a given trained ML or DL model can then be back-transformed to be spatially correlated by inverting the decorrelating transformation.  Importantly, our proposed decorrelation transform is computationally scalable because it relies only on nearest neighbor information so as to keep the rank of covariance matrices small and hence, applicable to big spatial data.

Section \ref{methods} motivates the decorrelating transformation from properties of multivariate Gaussian distributions and Vecchia approximations.  Section \ref{simStudies} demonstrates the decorrelation method and compares its performance to other spatial and non-spatial (baseline) methods.  Section \ref{app} applies our decorrelation approach on a real data set of particulate matter content across the contiguous United States.  Finally, Section \ref{conc} concludes and discusses areas of future research.

\section{Methodology}\label{methods}
\subsection{Background}
Let $\bm{Y} = (Y(\bell_1),\dots,Y(\bell_n))'$ be a response variable measured at the locations $\bell_1,\dots,\bell_n$ where $\bell_i \in \mathcal{D} \subset \mathbb{R}^d$ and $\mathcal{D}$ is the spatial domain and $d$ is an integer (typically, $d=2$ in spatial statistics such that $\bell_i = (\ell_{i1},\ell_{i2})$ would correspond to longitude and latitude coordinates).  Further let $\bm{x}(\bell) = (x_0(\bell),\dots,x_{P}(\bell))'$ be a vector of features (covariates) where we assume that $x_0(\bell) \equiv 1$ is an intercept term (also commonly referred to as a bias term).  As motivation for our methods, we assume $Y(\bell)$ follows a GP with linear mean $\mu(\bell) = \bm{x}'(\bell)\bm{\beta}$ where $\bm{\beta} = (\beta_0,\dots,\beta_P)'$ is a vector of coefficients such that
\begin{align}
    \bm{Y} &\sim \mathcal{N}_n(\bm{X}\bm{\beta}, \sigma^2\bm{R}) \label{linGP}
\end{align}
where $\mathcal{N}_n$ is the $n-$dimensional Gaussian distribution, $\bm{X}$ is the $n \times (P+1)$ design matrix of features (with a leading column of 1's), $\sigma^2$ is a common variance (also referred to as the sill in spatial statistics terminology) and $\bm{R} = \{\rho_{ij}\}$ is the $n\times n$ spatial correlation matrix.  For exposition (although not necessary for our proposed methods), we assume a stationary correlation (kernel) function such that
\begin{align}
    \rho_{ij} &= \begin{cases} 1 &\text{if } i = j \\
    (1-\omega)\rho(\|\bm{\ell}_i-\bm{\ell}_j\| \mid \bm{\phi}) & \text{if } i \neq j
    \end{cases}\label{covFX}
\end{align}
where $\omega \in [0,1]$ is a nugget term \citep[see][Chapter 2 for details]{banerjee2014hierarchical} and $\rho(d \mid \bm{\phi})$ is a stationary correlation function governed by parameters $\bm{\phi}$ (e.g.\ range and smoothness parameters).

The factorization of the $n\times n$ matrix $\bm{R}$, where $n$ is the number of observations, makes model \eqref{linGP} computationally infeasible for even moderately sized data sets.  To develop a more computationally feasible approach for spatial data that can subsequently be extended to machine and deep learning models, recall that any multivariate probability density function (PDF) can be factored as a series of univariate conditional distributions.  In our consideration, the multivariate Gaussian distribution in \eqref{linGP} can be equivalently factored as a series of univariate Gaussian conditional distributions such that,
\begin{align}
    \mathcal{N}_n(\bm{Y} \mid \bm{X}\bm{\beta}, \sigma^2\bm{R}) &\equiv \mathcal{N}_1(Y(\bell_1) \mid \mu_1, \sigma^2) \prod_{i=2}^n \mathcal{N}_1(Y(\bell_i) \mid \mu_i, \sigma^2v_i)\label{condFac}
\end{align}
where $\mathcal{N}_1(y \mid m, v)$ is the univariate Gaussian distribution for $y$ with mean $m$ and variance $v$.  While not explicitly shown in \eqref{condFac}, the distributions in the product are conditional distributions on previous data points $\{Y(\bell_1),\dots,Y(\bell_{i-1})\}$ through the mean and variance terms which are, according to properties of the multivariate normal distribution,
\begin{align}
    \mu_i &= \begin{cases} \bm{x}'(\bm{s}_1)\bm{\beta} &\text{if } i = 1 \\
    \bm{x}'(\bm{s}_i)\bm{\beta} + \bm{R}(i, \mathcal{L}_i)\bm{R}^{-1}(\mathcal{L}_i, \mathcal{L}_i)(\bm{Y}_{\mathcal{L}_i}-\bm{X}_{\mathcal{L}_i}\bm{\beta}) & \text{if } i > 1
    \end{cases} \label{mui} \\
    v_i &= \begin{cases} 1 & \text{if } i = 1 \\
 1- \bm{R}(i, \mathcal{L}_i)\bm{R}^{-1}(\mathcal{L}_i, \mathcal{L}_i)\bm{R}(\mathcal{L}_i,i) & \text{if } i > 1
 \end{cases}
 \label{vi}
\end{align}
where $\mathcal{L}_i = \{1,\dots,i-1\}$ is the set of indices of preceding points, $\bm{Y}_{\mathcal{A}}$ is the set of $Y(\bm{s})$ corresponding to indices in the $\mathcal{A}$, $\bm{X}_{\mathcal{A}}$ are the rows of $\bm{X}$ corresponding to $\mathcal{A}$ and $\bm{R}(\mathcal{A},\mathcal{B})$ is the submatrix of $\bm{R}$ with rows indexed in $\mathcal{A}$ and columns in $\mathcal{B}$.

Importantly, the factorization in \eqref{condFac} is still computationally demanding because the forms for $\mu_i$ and $v_i$ in \eqref{mui} and \eqref{vi} require dealing with large matrices through $\bm{R}(\mathcal{L}_i, \mathcal{L}_i)$ because the cardinality of $\mathcal{L}_i$ grows as $i \rightarrow n$.  Hence, for computational feasibility, we adopt the Vecchia process approximation framework \citep[see][]{datta2016hierarchical, datta2016cholesky, katzfuss2021general} by replacing $\mathcal{L}_i$ with a conditioning set $\mathcal{C}_i$ which we define to be the set of the at most $C$ nearest neighbors of $\bm{\ell}_i$ in terms of Euclidean distance.  In this way, the cardinality of $\mathcal{C}_i$ is at most $C$ so that $\bm{R}(\mathcal{C}_i, \mathcal{C}_i)$ is also at most $C \times C$ and can be dealt with computationally.  This Vecchia approximation relies on the assumption that all the information about $Y(\bm{\ell}_i)$ in the conditional distribution $\mathcal{N}_i(Y(\bm{\ell}_i) \mid \mu_i, \sigma^2 v_i)$ can be adequately summarized by the $C$ nearest neighbors to $Y(\bm{s}_i)$ among $Y(\bm{\ell}_1),\dots,Y(\bm{\ell}_{i-1})$.  

In using the Vecchia approximation, we have to consider (i) the choice of $C$ and (ii) the ordering of the observations.  Early investigations into Vecchia approximations by \citet{datta2016hierarchical} found that $C \approx 20$ was typically sufficient for isotropic spatial processes.  Subsequent research by \citet{katzfuss2021general} showed that often $C \approx 10$ was still sufficient.  In early stages of this research, we found that $C >20$ generally captured the spatial correlation.  Hence, here, we use $C=30$ as this goes beyond these guidelines while maintaining computational efficiency.  In regards to the ordering,  we note that \cite{katzfuss2021general} discuss the impact of observation ordering on this Vecchia assumption and recommend certain orderings of the observations to obtain better approximations.  For purposes of this research, follow their max-min suggested ordering.

\subsection{Proposed Spatial Adjustment}  

Under the factorization in \eqref{condFac}, consider the transformation
\begin{align}
    \tilde{Y}(\bell_i) &= \begin{cases} Y(\bell_i) & \text{ if } i = 1 \\
    v_i^{-1/2}\left(Y(\bell_i) - \bm{R}(i, \mathcal{C}_i)\bm{R}^{-1}(\mathcal{C}_i, \mathcal{C}_i)\bm{Y}_{\mathcal{C}_i}\right) & \text{ if } i>1.
    \end{cases}
    \label{transformation}
\end{align}
Under this transformation, if $Y(\bell_i)$ is Gaussian, $\tilde{Y}(\bell_i)$ will also be Gaussian with expectation
\begin{align}
    \mathbb{E}(\tilde{Y}(\bell_i) \mid \bm{Y}_{\mathcal{L}_i}) &= \tilde{\bm{x}}'_i(\bell_i)\bm{\beta} 
    \label{YtildeExp}
\end{align}
where
\begin{align}
    \tilde{\bm{x}}'(\bell_i) &= \begin{cases} \bm{x}'(\bell_i) & \text{ if } i = 1 \\
    v_i^{-1/2}\left(\bm{x}'(\bell_i) - \bm{R}(i, \mathcal{C}_i)\bm{R}^{-1}(\mathcal{C}_i, \mathcal{C}_i)\bm{X}_{\mathcal{C}_i}\right) & \text{ if } i>1.
    \end{cases} \label{xtrans}
\end{align}
and variance $\sigma^2$.  Specifically, under \eqref{transformation}, $\tilde{Y}(\bell_i) \overset{ind}{\sim} \mathcal{N}(\tilde{\bm{x}}_i(\bell_i)\bm{\beta}, \sigma^2)$ or, equivalently, $\tilde{\bm{Y}} \sim \mathcal{N}(\tilde{\bm{X}}\bm{\beta}, \sigma^2\bm{I})$ where $\tilde{\bm{Y}} = (\tilde{Y}(\bell_1),\dots,\tilde{Y}(\bell_n))'$ and $\bm{I}$ is an identity matrix.  Notably, \eqref{transformation} under a Gaussian assumption, is a spatial decorrelating transformation.

Given the above properties, we propose using \eqref{transformation} and \eqref{xtrans} as a decorrelating transform that can be done as a pre-processing step prior to machine and deep learning model fitting.  Importantly, while the transformation is motivated by properties of a multivariate Gaussian distribution, in ML and DL approaches we do not generally assume that $\bm{Y}$ follows a Gaussian distribution.  Rather, we are simply applying \eqref{transformation} to spatial data in an effort to reduce the impact of spatial correlation on model training.

Note the following key impacts of the transformation of $Y(\bell_i)$ to $\tilde{Y}(\bell_i)$ via \eqref{transformation} on ML or DL approaches.  First, because \eqref{transformation} theoretically removes the effect of the neighboring observations $\bm{Y}_{\mathcal{C}_i}$ on $Y(\bell_i)$, methods that resample the data (such as random forests) or use minibatching (such as neural networks) can be applied directly without regard to the spatial structure.  When correlated, subsets of $\bm{Y}$ cannot be considered because of the spatial correlation affecting the entire vector \citep[see][]{saha2023random}.  However, subsets or bootstrap samples of $\tilde{\bm{Y}}$ can be considered without regard for the spatial structure due to independence.

Second, the transformed matrix of feature variables (inputs) $\tilde{\bm{X}}$ forms a linear basis for $\tilde{\bm{Y}}$ if the data are Gaussian.  Hence, when using this transformation to move beyond linearity in ML and DL approaches, we propose using $\tilde{\bm{x}}(\bell)$ from \eqref{xtrans} as the appropriate input features into the ML and DL function rather than the original $\bm{x}(\bell)$. Importantly, because the untransformed $\bm{x}(\bell_i)$ \textit{includes} an intercept term (i.e.\ $x_0(\bell) \equiv 1$), after the transformation in \eqref{xtrans}, $\tilde{\bm{x}}(\bell)$ includes a transformed intercept term.  This transformed intercept term is important because, the conditioning set indexed by $\mathcal{C}_i$ varies with observation with the first observation being a marginal distribution as per Equation \eqref{condFac}.  Keeping the intercept term in the transformation of $\bm{x}(\bell_i)$ to $\tilde{\bm{x}}(\bell_i)$ accounts for this variability in the conditioning set and is needed for our approach in spite of intercepts not being traditionally included as features in machine or deep learning models. 

Third, by scaling by $v_i^{-1/2}$ all the $\tilde{Y}(\bm{\ell}_i)$ have a common variance $\sigma^2$. Under a common variance, common loss functions such as squared error loss are appropriate (rather than, e.g., weighted least squares). Hence, any machine or deep learning model can be fit in the standard way according to any chosen loss function.

After training the machine learning model on the transformed  data $\{\tilde{Y}(\bell_i), \tilde{\bm{x}}(\bell_i)\}_{i=1}^n$, predictions for a new response, $Y(\bu)$, at an unobserved location $\bu \in \mathcal{D}$ is obtained as follows.  First, let $\mathcal{C}_u \subset \{1,\dots,n\}$ denote the set of $C$ nearest neighbors to $\bu$ from among the training set locations $\bell_1,\dots,\bell_n$.  Given $\mathcal{C}_u$, define $\tilde{\bm{x}}(\bu)$ as in \eqref{xtrans} then input $\tilde{\bm{x}}(\bu)$ into the fitted machine learning model to get a prediction $\tilde{Y}^\star(\bu) = \hat{f}(\tilde{\bm{x}}(\bu))$.  We can then back transform the prediction $\tilde{Y}^\star(\bu)$ to a prediction for $Y(\bu)$ (i.e.,\ recorrelate the prediction with its spatial neighbors) via the inverse transformation
\begin{align}
    Y^\star(\bu) &= v^{1/2}\tilde{Y}^\star(\bu) + \bm{R}(\bu, \mathcal{C}_u)\bm{R}^{-1}(\mathcal{C}_u, \mathcal{C}_u)\bm{Y}_{\mathcal{C}_u}
    \label{backtrans}
\end{align}
where we use the notation $\bm{R}(\bu, \mathcal{N}_u)$ to denote $\text{Corr}(Y(\bu), \bm{Y}_{\mathcal{N}_u})$ under the correlation function in \eqref{covFX}.

In brief, our proposed method is as follows. First, inputs and outputs are transformed according to \eqref{transformation} and \eqref{xtrans}.  Second, the machine or deep learning model of choice is fit to the transformed data $\{(\tilde{Y}(\bell_i),\tilde{\bm{x}}(\bell_i))\}_{i=1}^n$.  Third, the model predicts the transformed $\tilde{Y}^\star(\bm{u}) = \hat{f}(\tilde{\bm{x}}(\bu))$ and the output is backtransformed via \eqref{backtrans}.  This algorithm is detailed as Algorithm \ref{alg:SpatialAdj}.

\begin{algorithm}[tb]
\caption{Spatial Adjustment for Machine and Deep Learning}\label{alg:SpatialAdj}
\begin{algorithmic}[1]
\State Choose spatial tuning parameters $\omega \in [0,1]$ and $\bm{\phi}$ as well as machine-specific tuning parameters $\bm{\theta}$.
\State Order observations according to the max-min criterion of \cite{katzfuss2021general}.
\State Determine nearest neighbor sets $\mathcal{L}_i$ for $i=1,\dots,n$.
\State Transform $Y(\bell_i)$ to $\tilde{Y}(\bell_i)$ via \eqref{transformation}.
\State Transform $\bm{x}(\bell_i)$ to $\tilde{\bm{x}}(\bell_i)$ via \eqref{xtrans}.
\State Fit machine or deep learning model $\hat{f}(\tilde{\bm{x}}(\bell))$ to transformed data $\{(\tilde{Y}(\bell_i),\tilde{\bm{x}}(\bell_i))\}$. 
\ForEach{prediction location $\bm{u} \in \mathcal{D}$}
    \State Determine nearest neighbor set of $\bm{u}$ ($\mathcal{L}_{\bm{u}}$) among training locations $\{\bell_1,\dots,\bell_n\}$
    \State Transform $\bm{x}(\bu)$ to $\tilde{\bm{x}}(\bu)$ via \eqref{xtrans}
    \State Obtain a prediction $\tilde{Y}^\star(\bu) = \hat{f}(\tilde{\bm{x}}(\bm{u}))$.
    \State Back transform prediction $\tilde{Y}^\star(\bu)$ to original scale prediction $Y^\star(\bu)$ according to \eqref{backtrans}.
    \State Output prediction $Y^\star(\bu)$.
\EndFor
\end{algorithmic}
\end{algorithm}

\subsection{Notes on Implementation}
The above adjustment for spatial correlation in the data are general and allows the user to fit any machine or deep learning model of choice.  However, there are a few details that are important to consider in the implementation which we briefly discuss here.  First, our proposed spatial adjustment is inherently different from \citet{saha2023random} and \citet{zhan2023neural} in that these approaches use a generalized squared error loss function $(\bm{Y}-\bm{f}(\bm{X}))'\bm{R}^{-1}(\bm{Y}-\bm{f}(\bm{X}))$ which is \textit{not} the same as an independent squared error loss function $(\tilde{\bm{Y}}-\bm{f}(\bm{\tilde{X}}))'(\tilde{\bm{Y}}-\bm{f}(\bm{\tilde{X}}))$ because the correlation matrix $\bm{R}$ doesn't commute into the function $f(\cdot)$ except in linear cases.  Consequently, the function $f(\bm{x}(\bell))$ learned under the two loss functions will \textit{not} be the same function.  Generally, this is not an issue as interpretation of $f(\bm{x}(\bell))$ is not of direct interest.  Further, fitting under independent squared error loss will be substantially faster than under the generalized squared error loss.

For brevity in exposition, in the above description we used the exponential correlation function given by \eqref{covFX}.  However, the spatial decorrelation transform in \eqref{transformation} is not dependent upon such a simplistic correlation structure.  In fact, any correlation structure, including anisotropic or nonstationary, could be used in \eqref{transformation}.  The transformation is only dependent upon knowing the parameters of the chosen correlation function which we now address.

In practice, Algorithm \ref{alg:SpatialAdj} requires knowing the parameters of the correlation function ($\omega$ and $\bm{\phi}$ in \eqref{covFX}) in addition to any tuning parameters of the chosen model. These spatial correlation parameters can be dealt with in two ways.  First, from a machine learning perspective, these parameters become extra tuning parameters that can be chosen via cross-validation.  For most correlation functions, these parameters have reasonable bounds (e.g.\ $\omega \in [0,1]$) so that a grid search approach could be used for tuning.  Likewise, Bayesian search algorithms could also be used \citep{wu2019hyperparameter,turner2021bayesian}. Or, second, these parameters could be estimated from the data using maximum likelihood estimation.  That is, a subset of $\bm{Y}$ could be randomly sampled and used to estimate $\omega$ and $\bm{\phi}$ to use in the algorithm.  In the simulation studies and applications below, we treat these correlation parameters as additional tuning parameters rather than estimating them from a subsample.  We do this because the amount of spatial correlation in the residuals depends on the type of mean function used (e.g.\ a nonlinear mean function will likely have a different correlation structure than a linear one because the mean function captures different features of the data).  Hence, by tuning these spatial parameters, we optimize the spatial correlation parameters to the type of mean function being fit.

% When predictions at multiple unobserved locations, say $\bu_1,\dots,\bu_K$, are needed (which is typically the case), two options are possible.  First, as we outline in Algorithm \ref{alg:SpatialAdj} above, prediction at each $\bu_k$ is done individually (and in parallel) by defining the nearest neighbor conditioning set $\mathcal{C}_{\bu}$ to be defined in terms of the training set $\{1,\dots,n\}$.  This approach is advocated by \cite{datta2016cholesky} and is computationally fast but has the disadvantage that it doesn't correlate the predictions $Y^\star(\bu_1),\dots,Y^\star(\bu_K)$ with each other but only with the observed data.  If the correlation between predictions $Y^\star(\bu_1),\dots,Y^\star(\bu_K)$ needs to be captured, a sequential approach can be taken wherein, $\bu_1$, $Y^\star(\bu_1)$ and $\bm{x}(\bu_1)$ could be included when generating a prediction at $\bu_2$ and so on.  However, we note that this sequential approach is computationally expensive and we found that it is not worth the effort in terms of predictive accuracy.  Hence, in this article, we use the approach in Algorithm \ref{alg:SpatialAdj}.

\section{Simulation Studies}\label{simStudies}
\subsection{Setup}
In this section, we demonstrate the performance of our spatial adjustment in 3 simulated scenarios.  In all scenarios, we simulate 50 datasets of size $n=50000$ where spatial locations $\bm{\ell}_i$ are distributed uniformly on the unit square for $i=1,\dots,n$ of which 40,000 observations were used for training and 10,000 used for testing.  In the first scenario, the response is simulated according to a standard linear regression model where $\bm{Y} \sim \mathcal{N}_n(\bm{X}\bm{\beta}, \sigma^2\bm{I})$ where $\bm{X}$ is a $n\times 10$ matrix of features where elements were simulated independently from a standard Gaussian distribution.  We drew the true $\bm{\beta}$ parameters independently from $\mathcal{N}(0,5^2)$ distribution and then randomly selected $J$ of the 10 covariates to have zero effect where $J$ followed a binomial distribution with 10 trials (corresponding to each of the 10 covariates) and success probability $0.5$.  Thus, the complete set of features contained features that did not relate to the response.  Finally, we set $\sigma^2 = 100$ which roughly corresponds to an expected coefficient of determination of 0.5.

The second simulated scenario is the similar to the first other than we let $\bm{Y} \sim \mathcal{N}_n(\bm{X}\bm{\beta},\sigma^2\bm{R})$ where $\bm{R}$ is a spatial correlation matrix constructed using the correlation function in \eqref{covFX}.  Specifically, we set $\rho(\cdot)$ as the exponential correlation with $\phi = 0.236$ and a nugget $\omega = 0.25$.  Setting $\phi=0.236$ corresponds to a spatial range (i.e., the distance at which the correlation decays to 0.05) of $\sqrt{2}/2$ which is half of the maximum spatial distance on the unit square (the spatial domain).

Finally, in the third scenario we simulate $\bm{Y} \sim \mathcal{N}_n(\bm{f}(\bm{X}), \sigma^2\bm{R})$ where $\bm{f}(\bm{X})$ is a non-linear function simulated from a zero-mean Gaussian process in $x_1$ and $x_2$.  That is $\bm{f} \sim \mathcal{N}_n(\bm{0}, \sigma^2_f\bm{M})$ where $\bm{M}$ is determined by \eqref{covFX} with a Matern correlation function in terms of $\|\bm{x}_i-\bm{x}_j\|$ with smoothness $2.1$, range $0.842$ and $\omega = 0$ (the value of $\phi$ was chosen so that the spatial range was half of the maximum distance).  By simulating $\bm{f}$ in this manner, the resulting relationship between the response and features is non-linear.

Each of the three simulation scenarios described above correspond to increasing complexity in the data.  In the first scenario, the relationship between the features $(\bm{X})$ and the response is linear and without spatial correlation.  We use this scenario to assess how well the spatial transformation in \eqref{transformation} works when no spatial correlation is present.  The second scenario only adds spatial correlation to the first scenario and can be used to assess the impact of accounting for spatial correlation in a simple relationship between the response and features.  Finally, the third scenario is the most complex as the relationship between the response and features is non-linear and there is spatial correlation among the response.

We employ linear models (LM), Bayesian additive regression trees (BART), single-layer perceptrons (SLP), boosting, random forests (RF) and $K$ neareast neighbors (KNN) both with and without the spatial adjustment proposed in \eqref{transformation}. For BART, we consider only the number of trees as a tuning parameter.  For SLP, we tune the number of hidden neurons, penalty parameters on the weights and the number of epochs while we fix the activiations as ReLU with no dropout.  For boosting, we tune the tree depth, number of trees and learning rate.  For RF, we tune the number of variables at each split and minimum number of observations per leaf but we fix the number of trees at 250.  Finally, for KNN, we tune the number of neighbors. Additionally, following \citet{georganos2021geographical}, we also fit a geographical RF (GeoRF) using spatial features as inputs.  Note that a non-spatial GeoRF is simply a RF. To tune the above algorithms, we use a grid search across 5 values of each tuning parameter and with finalized tuning parameters chosen to minimize the 5-fold cross-validation root mean square error (RMSE).

 Ideally, we would have also compared the spatial random forests of \citet{saha2023random} as well as the spatial neural network of \citet{zhan2023neural}.  However, implementing these methods on a single dataset took over 24 hours for the spatial random forest and 8.6 hours for the spatial neural networks.  Hence, not only fitting but tuning these methods was not computationally feasible.

\subsection{Results}
Table \ref{tab:CombinedSimulationResults} displays the RMSE for each method both with and without the spatial adjustment on the simulated data in each of the scenarios.   For Scenario 1 (linear, independent), all models exhibit similar performance. This equivalence is expected because of the simplicity of the simulation scenario.  However, these results are important to consider because they indicate that the spatial transformation method does not hinder the prediction performance when there is no spatial correlation in the data. That is, the spatial transformation can be tuned to account for the apparent lack of spatial correlation in the data.  In contrast, note that the RMSE under GeoRF is higher suggesting that adding spatial features when no spatial correlation is present can lead to a decrease in predictive accuracy.
\begin{table}[!tb]
  \centering
  \caption{Comparison of predictive (out-of-sample) RMSE for spatial and non-spatial models across the 3 simulation scenarios.  Note that the GeoRF method always includes a spatial aspect and, hence, does not have a corresponding non-spatial RMSE.  The proposed spatial adjustment improves the RMSE for the spatial scenarios while not impacting the RMSE for the case of spatial independence.}
  \label{tab:CombinedSimulationResults}
  
  \begin{tabular}{|c|c|c|c|c|c|c|}
    \hline
    \multirow{2}{*}{Model} & \multicolumn{2}{c|}{Independent Linear} & \multicolumn{2}{c|}{Spatial Linear} & \multicolumn{2}{c|}{Spatial Non-Linear} \\
    \cline{2-7}
     & Non-Spatial & Spatial & Non-Spatial & Spatial & Non-Spatial & Spatial \\
    \hline
    RF & 10.3 & 10.3 & 9.17 & 8.65 & 9.33 & 6.28 \\
    BART & 10.0 & 10.0 & 8.95 & 8.20 & 9.31 & 6.26 \\
    Boost & 10.2 & 10.2 & 9.12 & 8.29 & 9.41 & 6.27 \\
    KNN & 10.7 & 10.7 & 9.72 & 9.65 & 9.84 & 6.97 \\
    LM & 9.99 & 9.99 & 8.90 & 5.26 & 10.5 & 7.39 \\
    SLP & 10.1 & 10.1 & 8.99 & 8.87 & 9.63 & 6.67 \\
    GeoRF & NA & 11.31 & NA & 10.15 & NA & 9.57 \\
    \hline
  \end{tabular}
\end{table}
% \begin{figure}[tb]
%     \centering
%     \includegraphics[width=0.75\linewidth]{NewIndSim.jpeg}
%     \caption{Enter Caption}
%     \label{fig:IndependentSim}
% \end{figure}

Under Scenario 2 (linear, spatial), not surprisingly, the linear model that accounts for spatial correlation performed best (lowest RMSE) because this was the exact data generating model. However, note that the spatial versions of each of the algorithms outperform the non-spatial versions.  The degree of reduction ranged from a 40\% reduction in RMSE (linear) to only a 1\% reduction in RMSE (KNN).  Finally, GeoRF still came in last suggesting that adding the spatial features did not adequately capture the spatial correlation. These results suggest that accounting for spatial correlation in the model improved performance over non-spatial models.  However, as we will see below, the modest reduction in RMSE is likely due to the simplistic linear nature of the relationship between the features and the response.
% \begin{figure}[tb]
%     \centering
%     \includegraphics[width=0.75\linewidth]{NewLinSim.jpeg}
%     \caption{Enter Caption}
%     \label{fig:LinSim}
% \end{figure}

In Scenario 3 (non-linear, spatial), spatial models demonstrate superior performance compared to their non-spatial counterparts. The improvement seen by using spatial correlation ranges from a 33\% reduction (Boosting) to 29\% reduction (KNN).  This larger reduction in RMSE on non-linear data are due to appropriately accounting for the spatial nature of the data.  In the non-spatial settings, the models likely overfit spatial structure resulting in worse out-of-sample error.  
% \begin{figure}[htbp]
%     \centering
%     \includegraphics[width=0.75\linewidth]{NewNonLinSim.jpeg}
%     \caption{Enter Caption}
%     \label{fig:NonLinSim}
% \end{figure}

As a final comparison, we compare the computation time associated with each model.  Overall, the spatial adjustment for $n=50,000$ took 90 seconds on Apple M1 chip with 64 GB of memory.  After the spatial adjustment, LM took 4 seconds, RF 49 seconds, BART 88 seconds, boosting 23 seconds, KNN 38 seconds, SLP 12 seconds, and GeoRF 49 seconds.  By comparison, as mentioned above, the current R and python implementations for the spatial random forests of \citet{saha2023random} and the spatial neural networks of \citet{zhan2023neural} took more than 24 hours and 8.6 hours, respectively.

\section{Application}\label{app}
\begin{figure}[tb]
    \centering
    \includegraphics[scale=.45]{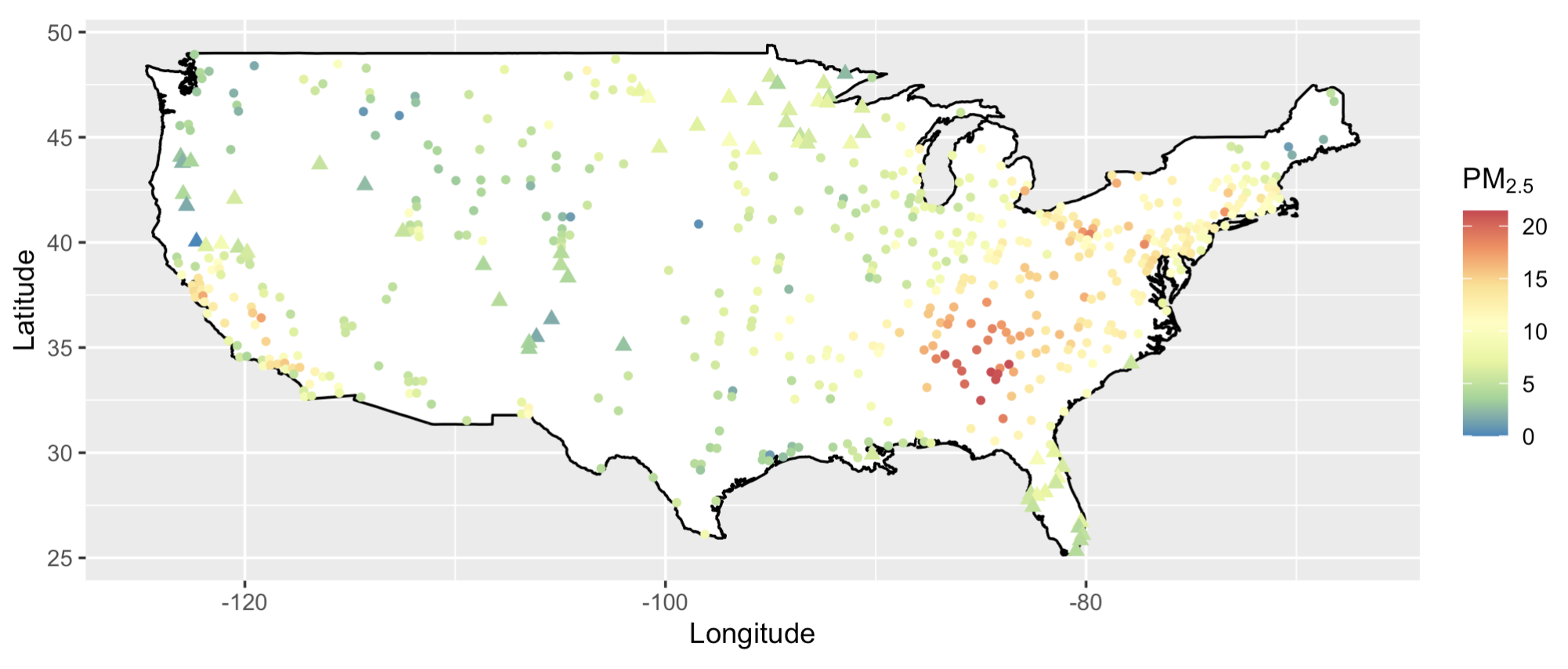}
    \caption{Raw PM$_{2.5}$ data along with an example train-validation split for assessing predictive performance.  The validation set is given by the triangle points.}
    \label{fig:rawPM}
\end{figure}

In this section, we demonstrate the performance of our spatial adjustments using an application in pollution monitoring.  Importantly, in contrast to the simulations above, this application is not generated from a Gaussian process.  Hence, using this application, we wish to see how the spatial adjustment improves ML and DL methods in this setting.  Specifically, we analyze data on particulate matter less than 2.5 micrometers (PM$_{2.5}$) in diameter taken from the environmental protection agencies (EPA) network of monitors.  We utilize the same PM$_{2.5}$ application as \citet{zhan2023neural}.  After cleaning, the data we consider consists of 593 measurements of PM$_{2.5}$ across the contiguous United States taken on June 5, 2019.  We further follow \citet{zhan2023neural} and split the data into train and validation sets according to a block-random split strategy which removes whole areas of data and is closer to a real-world scenario (see the Appendix S5.1 of \citealt{zhan2023neural} for details).  This split method resulted in 72 different train-validation splits with a split ratio of approximately 80\%-20\%.  The raw data along with one train-validation split is shown in Figure \ref{fig:rawPM}.

\begin{figure}[tb]
    \centering
    \includegraphics[scale=.23]{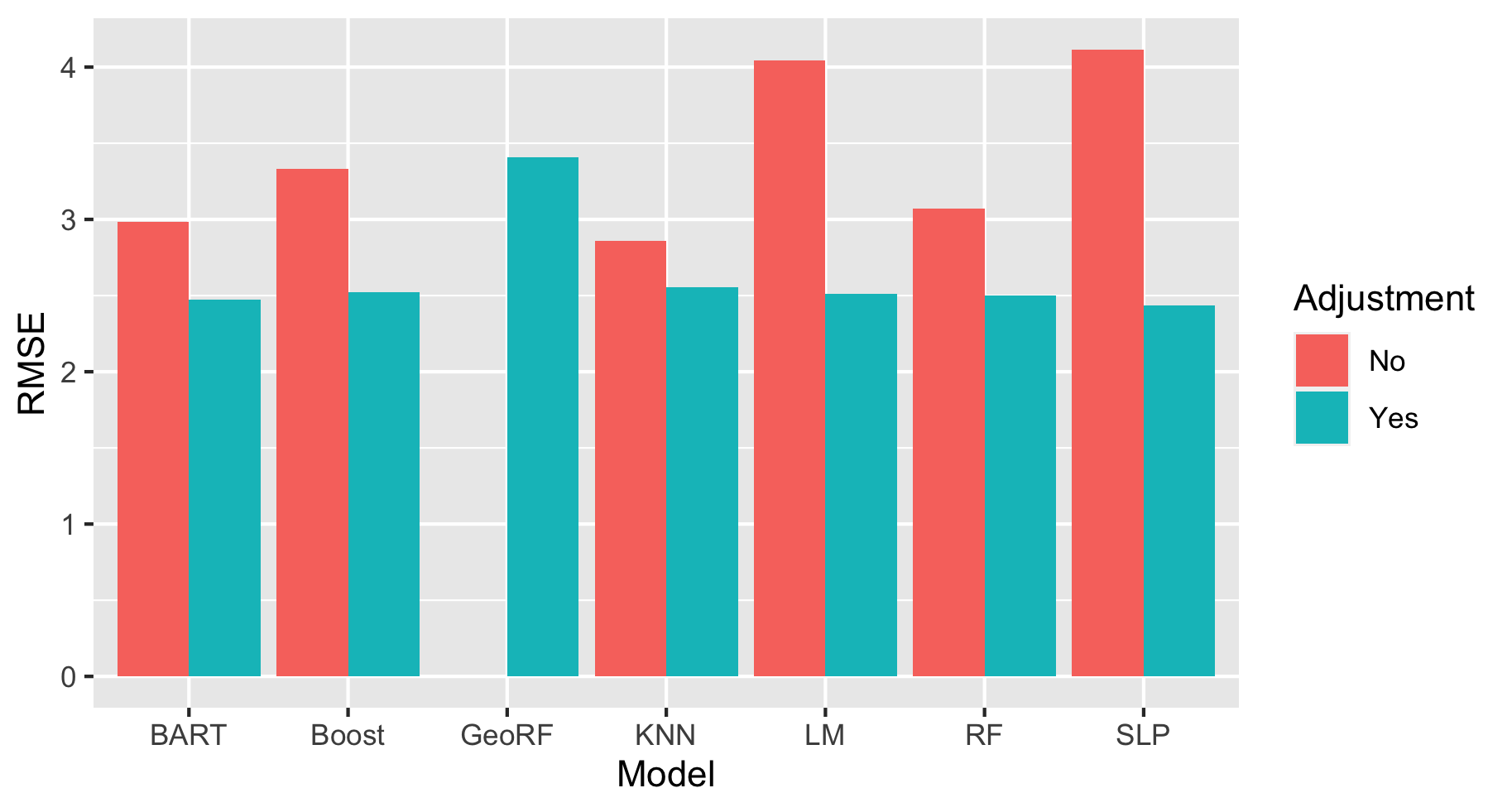}
    \caption{Median RMSE for all models both with and without the adjustment for spatial correlation for the PM$_{2.5}$ example.  Note that the spatial adjustment, in all models, decreases the RMSE.  The GeoRF model always includes spatial features and, hence, does not have a ``non-spatial'' equivalent.}
    \label{fig:PMRMSE}
\end{figure}

Figure \ref{fig:PMRMSE} displays the median root mean square error (RMSE) across all the train-validation splits using block-random splitting.  Notably, for all models, inclusion of the spatial adjustment decreases RMSE suggesting that accounting for spatial correlation improves the predictive ability of all models when spatial correlation is present.  Interestingly, there is little difference between the models if spatial correlation is accounted for suggesting that spatial correlation can adapt to the type of model being fit to improve predictive ability.

\begin{figure}[tb]
    \centering
    \includegraphics[scale=.55]{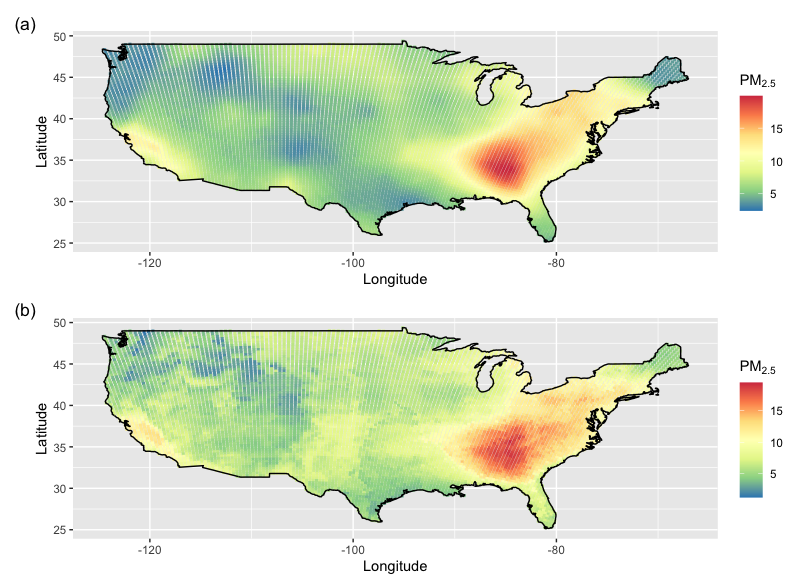}
    \caption{Comparison of predictions from BART model (a) using and (b) not using the spatial adjustment. The spatially-adjusted BART model exhibits smoother predictions than the non-spatial BART model.}
    \label{fig:PredComp}
\end{figure}

To understand further the impact of accounting for spatial correlation, Figure \ref{fig:PredComp} compares the predictions from the BART model both using and not using the spatial adjustment (we show results from BART here for illustration but all other models had similar results).  Clearly, using the spatial adjustment leads to smoother spatial predictions.  Even without the spatial adjustment, BART can capture some of the spatial structure. However, plain BART had a higher RMSE suggesting that BART may overfit the training data.

\section{Conclusions}\label{conc}
We present herein a spatial adjustment for using machine and deep learning methods on spatial data.  Intuitively, the method works by performing a spatial decorrelation transform to the data prior to model fitting.  Predictions from the fitted model are then backtransformed via the inverse transformation.  The method is able to be tuned to adjust to the amount of spatial correlation in the data and is computationally scalable so that it can be applied to large spatial datasets.  Via both simulated and real data applications, we demonstrated the added benefit of accounting for spatial correlation in common machine learning algorithms.

The examples used here were all done assuming a stationary spatial correlation structure (the most commonly used in applications).  While the decorrelation transform in \eqref{transformation} can presumably be done with any correlation function $\rho(\cdot)$, the application of more complex correlation structures such as anisotropic or non-stationary functions is an open area of research.  Admittedly, more complex spatial correlation structures often involve more parameters than only a spatial range and nugget term.  Hence, using the methods described here for these more complicated structures would require estimation (or tuning) of all the correlation function parameters.

Notably, the decorrelating transform \eqref{transformation} is only valid for quantitative data (note that the theoretical foundation was laid assuming a Gaussian assumption).  Obviously count, Bernoulli, multinomial or other data types can also exhibit spatial correlation.  In future research, we plan to address the issue of non-quantitative spatial data.

Given the breadth of ML and DL approaches, we are not able to do an exhaustive demonstration of the spatial adjustment in all approaches.  Particularly with DL methods, we only demonstrated how the spatial adjustment works in conjunction with multi-layer perceptrons.  As open questions, we need to consider how such a spatial adjustment may work with graphical or convolutional neural networks.

Finally, we note that we did not explore using our approach in conjunction with uncertainty quantification (UQ) for machine and deep learning.  Under our methods, the approach of, say,  \cite{zhang2019random} for random forests or any of the UQ methods for deep learning surveyed in \cite{gawlikowski2021survey} could be used to obtain $\tilde{Y}^\star_{\text{lwr}}(\bu)$, a lower bound, and $\tilde{Y}^\star_{\text{upr}}(\bu)$, an upper bound, on the decorrelated data which can then be backtransformed via \eqref{backtrans} to $Y^\star_{\text{lwr}}(\bu)$ and $Y^\star_{\text{upr}}(\bu)$. While this is a straightforward approach, what remains to be investigated is if such an approach maintains appropriate coverage under the spatial transform presented here.

\section*{Acknowledgements}
This material is based upon work supported by the National Aeronautics and Space Administration under Grant/Contract/Agreement No. 10053957-01 and by the National Science Foundation under Grant No. 2053188.  

\section*{Data Availability Statement}
Both R and Python functions to perform the spatial decorrelation transform are available on the authors GitHub page at \url{https://github.com/amillane/spatialtransform}.

% Manual newpage inserted to improve layout of sample file - not
% needed in general before appendices/bibliography.

\vskip 0.2in
\bibliographystyle{apalike}
\bibliography{mainDoc.bib}

\end{document}